\documentclass[twocolumn,floatfix]{revtex4}
\usepackage{epsf,graphicx}
\usepackage{amsmath,amssymb,amsfonts}
\newcommand{\vecbm}[1]{\mbox{\boldmath#1}}

\newcommand{\nvec}[1]{\stackrel{\rightarrow}{#1}}
\newcommand{\goo}{\,\raisebox{-.5ex}{$\stackrel{>}{\scriptstyle\sim}$}\,}
\newcommand{\stdef} {\stackrel{\mbox{\scriptsize def}}{=}}

\begin{document}
\title{Heat can flow from cold to hot in
\\ Microcanonical Thermodynamics of finite systems\\
The microscopic origin of condensation and phase separations}
\today
\author{D.H.E. Gross}
\affiliation{Hahn-Meitner Institute and Freie Universit{\"a}t Berlin,\\
Fachbereich Physik.\\ Glienickerstr. 100\\ 14109 Berlin, Germany}
\email{gross@hmi.de} \homepage{http://www.hmi.de/people/gross/ }
\begin{abstract}
Microcanonical Thermodynamics \cite{gross174} allows the application of
Statistical Mechanics on one hand to closed finite and even small systems
and on the other to the largest, self-gravitating ones. However, one has to
reconsider the fundamental principles of Statistical Mechanics especially
its key quantity, entropy. Whereas in conventional Thermostatistics the
homogeneity and extensivity of the system and the concavity of its entropy
$S(E)$ are central conditions, these fail for the systems considered here.
E.g. at phase separation the entropy $S(E)$ is necessarily convex to make
$e^{S(E)-E/T}$ bimodal in $E$ (the two coexisting phases). This is so even
for normal macroscopic systems with short-range coupling. As
inhomogeneities and surface effects in particular cannot be scaled away,
one has to be careful with the standard arguments of splitting a system
into two or bringing two systems into thermal contact with energy or
particle exchange \cite{wehrl78}: Not only the volume part of the entropy
must be considered. The addition of any other external constraint like a
dividing surface reduces the entropy. As will be shown here, when removing
such constraints in regions of a negative heat capacity, the system may
even relax under a flow of heat (energy) {\em against the temperature
slope}. Thus Clausius formulation of the Second Law: "Heat always flows
from hot to cold" can be violated. {\em Temperature is not a necessary or
fundamental control parameter of Thermostatistics}. However, the Second Law
is still satisfied and the total Boltzmann-entropy keeps constant or rises
when constraints are removed. In the final sections of this paper the
general microscopic mechanism leading to condensation and to the convexity
of the microcanonical entropy $S(E)$ at phase separation is sketched. Also
the microscopic conditions for the existence or non-existence of a critical
end-point of the phase-separation are discussed. This is explained for the
liquid--gas and the solid--liquid transition.
\end{abstract}
\maketitle

\section{What is entropy?}
Entropy ($S$) is {\em the} key quantity of all Thermostatistics
and Thermodynamics. Therefore, its proper understanding is
essential. In the literature this is sometimes clouded by the
frequent use of the thermodynamic limit and/or the use of the
Boltzmann-Gibbs canonical statistics. In what follows, I will try
to define it with minimum bias.

Boltzmann's epitaph reads
\begin{equation}
\fbox{\fbox{\vecbm{S=k*lnW}}}\label{boltzmann0}
\end{equation} with
\begin{equation}
W(E,N,{\cal V})=
\int{\frac{d^{3N}\nvec{p}\;d^{3N}\nvec{q}}{N!(2\pi\hbar)^{3N}}
\epsilon_0\;\delta(E-H\{\nvec{q},\nvec{p}\})}\label{boltzmann}
\end{equation} in semi-classical approximation. $E$ is the total energy, $N$ is
the number of particles and ${\cal V}$ the volume. Or, more
appropriate for a finite quantum-mechanical system:
\begin{eqnarray} W(E,N,{\cal V})&=& Tr[\mathcal{P}_E]\label{quantumS}\\
&=&\sum{\scriptsize\begin{array}{ll}\mbox{all eigenstates n of H with given
N,${\cal V}$,}\\\mbox{and } E-\epsilon_0/2<E_n\le E+\epsilon_0/2\nonumber
\end{array}}
\end{eqnarray}
and $\epsilon_0\approx$ the macroscopic energy resolution.  This is still
up to day the deepest, most fundamental, and most simple definition of
entropy. There is no need of the thermodynamic limit, no need of concavity,
extensivity and homogeneity.  In its semi-classical approximation,
eq.(\ref{boltzmann}), $W(E,N,{\cal V},\cdots)$ simply measures the area of
the sub-manifold of points in the $6N$-dimensional phase-space
($\Gamma$-space) with prescribed energy $E$, particle number $N$, volume
${\cal V}$, and some other time invariant constraints which are here
suppressed for simplicity. Because it was Planck who coined it in this
mathematical form, I will call it the Boltzmann-Planck principle.

There are various reviews on the mathematical foundation of Statistical
Mechanics e.g. the detailed and instructive article by Alfred Wehrl
\cite{wehrl78}. He shows how the Boltzmann-Planck formula (1), with (3) can
be generalized to the famous definition of entropy in Quantum Mechanics by
von Neumann\cite{neumann27}
\begin{equation}
S=-Tr[\rho\ln(\rho)],\label{neumann}
\end{equation}addressing general (non projector like) densities $\rho$.

Wehrl discusses the conventional canonical Boltzmann-Gibbs statistics where
all constraints are only fixed to their mean, allowing for free
fluctuations. He points to the many serious complications of this
definition. However, in the case of {\em conserved} observables we know
more than their mean. We know these quantities sharply. In {\em
Microcanonical }Thermodynamics we don't need von Neumann's definition
(\ref{neumann}) and can work on the level of Boltzmann-Planck's original
definition of $S$ by eqs.(\ref{boltzmann0} and \ref{quantumS}). We thus
explore Statistical Mechanics and entropy on its most fundamental level.
This has the great advantage that the axiomatic level is extremely simple.
Because it does not demand scaling or extensivity, it can further be
applied to the much wider group of non-extensive systems from nuclei to
galaxies \cite{gross207} and address the original object of Thermodynamics
for which it was invented some 150 years ago: phase separations.

The Boltzmann-Planck formula has a direct and simple but deep
physical interpretation: {\em $W$ or $S$ are the measure of our
ignorance about the complete set of initial values for all $6N$
microscopic degrees of freedom which are needed to specify the
$N$-body system unambiguously} \cite{kilpatrick67}. Usually we
only know the few time-independent control parameters $E,N,{\cal
V},\cdots$, of the system (conserved or very slowly varying) but
have no control of the other fast changing degrees of freedom. If
the information would be complete, i.e. all $6N$-degrees of
freedom sharply known at some time $t_0$, $W$ would be a single
cell of size $(2\pi\hbar)^{3N}$ in the $6N$-dim. phase space, and
$S$ would be $0$.

However, many initial values are consistent with the same values of
$E,N,{\cal V}, \cdots $, (redundancy of control parameters). All the other
degrees of freedom vary with time, usually fast, and are normally not under
control. The manifold of all these points in the $6N$-dim. phase space is
the microcanonical ensemble. It has a well defined geometrical size $W$ and
due to eq.(\ref{boltzmann0}) a non-vanishing entropy $S(E,N,{\cal
V},\cdots)$. The dependence of $S(E,N,{\cal V},\cdots)$ on its arguments
gives the complete thermostatics and equilibrium thermodynamics.

It is further clear that the Hamiltonian (Liouvillean) dynamics of the
system cannot create the missing information about the initial values. I.e.
the entropy $S(E,N,{\cal V},\cdots)$ cannot decrease. As is further worked
out in \cite{gross183} and more recently in \cite{gross207} the inherent
finite resolution of the macroscopic description implies usually an {\em
increase} of $W$ or $S$ with time when a constraint is removed (Second Law
of Thermodynamics).

It should be emphasized that this proper {\em statistical} definition of
the entropy $S(E,N,{\cal V},\cdots)$ characterizes the {\em whole}
microcanonical ensemble \cite{kilpatrick67}. It measures the total number
of the {\em possible} states of the system under the information given.
Eqs. (\ref{boltzmann},\ref{quantumS}) are in clear contrast to the
definition suggested by \cite{hertz10a,berdichevsky97,muenster69,rugh01}:
\begin{eqnarray}
W_{bulk}(E)&=&\int{\frac{d^{3N}\nvec{p}\;d^{3N}\nvec{q}}{N!(2\pi\hbar)^{3N}}
\Theta(E-H\{\nvec{q},\nvec{p}\})}\label{boltzmann*}\\\nonumber\\
&=&\sum{\scriptsize\begin{array} {ll}\mbox{all eigenstates n of H
with given N,${\cal V}$,}\\\mbox{and } 0\le E_n\le
E\end{array}}\label{quantumS*}\\ &=&e^{S_{bulk}(E,N,{\cal V})}
\label{sbulk}\end{eqnarray} Our {\em statistical} definition
(\ref{boltzmann},\ref{quantumS}) called $S_{surf}$ in
\cite{rugh01} has a priori not much to do with the adiabatic
invariants of a {\em single} trajectory in the $6N$-dim. phase
space.

As we will see later in section (\ref{chsplit}) the entropy $S(E)$ for
non-extensive systems can be convex in contrast to the assumption made by
Hertz \cite{hertz10a}. Then the temperature or the mean kinetic energy per
particle do {\em not} control the energy flow during the equilibration of
two systems in thermal contact as was the crucial argument of Hertz
\cite{hertz10a} to introduce $S_{bulk}$,(\ref{boltzmann*}-\ref{sbulk}). In
contrast, {\em the Boltzmann entropy eq.(\ref{boltzmann}) does still
control the direction of the energy-flow under equilibration} even in these
somewhat astonishing and unconventional though ubiquitous situations. This
clearly underlines the role of the  Boltzmann entropy (\ref{boltzmann}) as
the fundamental quantity and not the temperature for statistical mechanics.
I am fully aware that this contradicts opposite statements given by
\cite{hertz10a,rugh01,berdichevsky97}.

Moreover, the definition of the bulk entropy $S_{bulk}(E,N,{\cal V})$
(\ref{boltzmann*}-\ref{sbulk}) contains also, energetically, inaccessible
states with less energy, which are clearly excluded by our knowledge of the
total energy.  {\em Thus, $S_{bulk}$ does not represent the redundancy, or
ignorance, of the available information}. Therefore it is not acceptable by
conceptual reasons. This is a clear warning that one should not follow some
formal definitions and forget the original clear meaning of entropy and
statistics as is often done in the literature. Of course, for macroscopic
systems the difference between the two alternative definitions of the
entropy eqs.(\ref{boltzmann}-\ref{quantumS}) and
eqs.(\ref{boltzmann*},\ref{sbulk}) disappears.
\section{The Zero'th Law in conventional extensive Thermodynamics}
In conventional (extensive) thermodynamics thermal equilibrium of two
systems (1 \& 2) is established by bringing them into thermal contact which
allows free energy exchange\footnote{This and the next (III.) section
discuss mainly systems that have no other macroscopic (extensive) control
parameter besides the energy. E.g. the particle density is not changed and
there are no chemical reactions.}. Equilibrium is established when the
total entropy
\begin{equation}
S_{total}(E,E_1)=S_1(E_1)+S_2(E-E_1)\label{eq1}
\end{equation}
is maximal:
\begin{equation}
dS_{total}(E,E_1)|_E=dS_1(E_1)+dS_2(E-E_1)=0\label{eq2}.
\end{equation}
Under an energy flux $\Delta E_{2\to 1}$ from $2\to 1$ the total
entropy changes to lowest order in $\Delta E$ by
\begin{eqnarray}
\Delta S_{total}|_E&=&(\beta_1-\beta_2)\Delta E_{2\to 1}\\
\beta&=&dS/dE=\frac{1}{T}
\end{eqnarray}
Consequently, a maximum of $S_{total}(E=E_1+E_2,E_1)|_E$ will be approached
when
\begin{equation}
 \mbox{sign}(\Delta S_{total})=\mbox{sign}(T_2-T_1)\mbox{sign}(\Delta E_{2\to 1})>0
\end{equation}
From here Clausius' first formulation of the Second Law follows: "Heat
always flows from hot to cold". Essential for this conclusion is the {\em
additivity} of $S$ under the split (eq.\ref{eq1}). There are no
correlations, which are destroyed when an extensive system is split.
Temperature is an appropriate control parameter for extensive systems.
\section{No phase separation without a convex, non-extensive $S(E)$\label{chsplit}}
Since the beginning of Thermodynamics in the middle of the 19.century its
main motivation was the description of steam engines and the liquid to gas
transition of water. Here water prefers to become inhomogeneous and develop
a separation of the gas phase from the liquid, i.e. water boils. As
conventional canonical statistics works only for {\em homogeneous,
infinite} systems, phase separation remains outside of standard
Boltzmann-Gibbs thermo-statistics.

The weight $e^{S(E)-E/T}$ of the configurations with energy E in the
definition of the canonical partition sum
\begin{equation}
Z(T)=\int_0^\infty{e^{S(E)-E/T}dE}\label{canonicweight}
\end{equation} becomes here {\em bimodal}, at the transition temperature it has
two peaks, the liquid and the gas configurations which are separated in
energy by the latent heat. Consequently $S(E)$ must be convex and the
weight in (\ref{canonicweight}) has a minimum between the two pure phases.
Of course, the minimum can only be seen in the microcanonical ensemble
where the energy is controlled and its fluctuations forbidden. Otherwise,
the system would fluctuate between the two pure phases by an, for
macroscopic systems even macroscopic, energy $\Delta E\sim E_{lat}$ of the
order of the latent heat. I.e. {\em the convexity of $S(E)$ is the generic
signal of a phase transition of first order} and of
phase-separation\cite{gross174}. It is amusing that this fact that is
essential for the original purpose of Thermodynamics to describe steam
engines was never treated correctly in the past 150 years.

\begin{figure}[h]
\includegraphics*[bb = 39 80 454 292, angle=-90, width=6 cm, clip=true]
{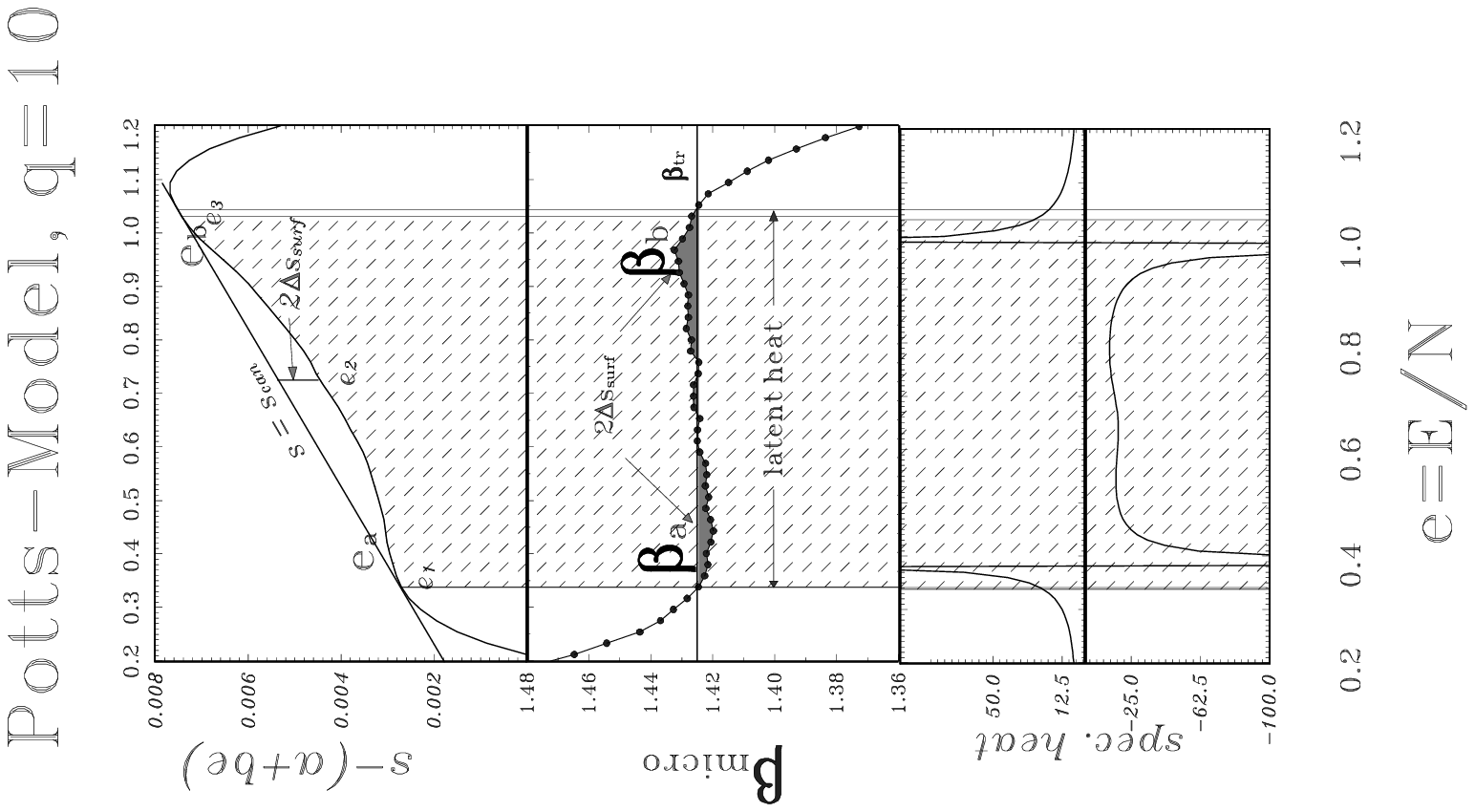}\caption{Ferromagnetic Potts model ($q=10$) on a $50*50$-lattice
with periodic boundary conditions in the region of phase separation. At the
energy $e_1$ per lattice point the system is in the pure ordered phase, at
$e_3$ in the pure disordered phase. At $e_a$ little above $e_1$ the
temperature $T_a=1/\beta$ is higher than $T_2$ and even more than $T_b$ at
$e_b$ a little below $e_3$.  At $e_a$ the system separates into a few
bubbles of disordered phase embedded in the ordered phase or at $e_b$ into
a few droplets of ordered phase within the disordered one. If we combine
two equal systems: one with the energy per lattice site $e_a=e_1+\Delta e$
and at the temperature $T_a$, the other with the energy $e_b=e_3-\Delta e$
and at the temperature $T_b<T_a$, and allowing for free energy exchange,
then the systems will equilibrize at energy $e_2$ with a {\em rise} of its
entropy. The temperature $T_a$ drops (cooling) and energy (heat) flows (on
average) from $b\to a$. I.e.: {\em Heat flows from cold to hot! Thus, the
Clausius formulation of the Second Law is violated}. This is well known for
self-gravitating systems. However, this is not a peculiarity of only
gravitating systems! It is the generic situation at phase separations
within classical thermodynamics even for systems with short-range coupling
and has nothing to do with long-range interactions.\label{figure 1}}
\end{figure}

The ferromagnetic Potts-model illuminates in a most simple example the
occurrence of a convex intruder in $S(E)$ which induces a backbending
caloric curve $T(E)=(\partial S/\partial E)^{-1}$ with a decrease of the
temperature $T(E)$ with rising energy \cite{gross150}\footnote{This
backbending of $T(e)$ has nothing to do with the periodic boundary
condition used here, as claimed erroneously by Moretto et.al
\cite{moretto02}. Many realistic microcanonical calculations for nuclear
fragmentation as also for atomic-cluster fragmentation, c.f.
\cite{gross95,gross174}, show the backbending {\em without using periodic
boundaries}. These are verified by numerous experimental data in nuclear
physics c.f. \cite{dAgostino00,borderie02} and cluster physics
\cite{schmidt01,schmidt02}. The errors of the above paper by Moretto et.
al. are commented in more detail also in \cite{gross196,gulminelli03}}. A
typical plot of $s(e,N)=S(E=Ne)/N$ in the region of phase separation is
shown in fig(\ref{figure 1}). Section \ref{convex} discusses the general
microscopic reasons for the convexity.

This has far reaching consequences which are crucial for the fundamental
understanding of thermo-statistics and Thermodynamics: Let us split the
system of figure (\ref{figure 1}) into two pieces $a$ \& $b$ by a dividing
surface, with half the number of particles each. The dividing surface is
purely geometrical. It exists only as long as the two pieces can be
distinguished by their different energy/particle $e_a$ and $e_b$.
Constraining the energy-difference $e_b-e_a=\Delta e$ between the two,
reduces the number of free, unconstrained degrees of freedom and {\em
reduces} the entropy by $-2\Delta S_{surf-corr.}$. (Moreover, if the effect
of the new surface would also be to cut some bonds: before the split there
were configurations with attractive interactions across the surface which
are interrupted by the division, their energy shifts upwards outside the
permitted band-width $\epsilon_0$, and thrown out of the partition sum
(\ref{quantumS}). I.e. the entropy will be further reduced by the split.)

If the constraint on the difference $e_b-e_a$ is fully relaxed and
$e_b-e_a$ can fluctuate freely at fixed $e_2=(e_a+e_b)/2$, the dividing
surface is assumed to have no further physical effect on the system.

For an {\em extensive} system [$S(E,N)=N  s(e=E/N)=2S(E/2,N/2)$]. One would
argue as follows: The combination of two pieces of $N/2$ particles each,
one at $e_a=e_2-\Delta e/2$ and a second at $e_b=e_2+\Delta e/2$, must lead
to $S(E_2,N)\ge S(E_a/2,N/2)+S(E_b/2,N/2)$, the simple algebraic sum of the
individual entropies because by combining the two pieces one normally
looses information. This, however, is equal to $[S(E_a,N)+S(E_b,N)]/2$,
thus $S(E_2,N)\ge[S(E_a,N)+S(E_b,N)]/2$. I.e. {\em the entropy $S(E,N)$ of
an extensive system is necessarily concave}.

For a {\em non-extensive} system we have in general $S(E,N)\ge 2S(E/2,N/2)$
because again two separated, closed pieces have more information than their
unification. Now, if $E_2$ is the point of maximum positive curvature of
$S(E,N)$ (convexity = upwards concave like $y=x^2$) we have $S(E_2,N) \le
[S(E_a,N)+S(E_b,N)]/2$ like in fig.(\ref{figure 1}). However, the r.h.s. is
larger than $S(E_a/2,N/2)+S(E_b/2,N/2)$. I.e. even though $S(E,N)$ is
convex at const. $N$, the unification of the pieces with $E_a/2,N/2$ and
$E_b/2,N/2$ can still lead to a {\em larger} entropy $S(E_2,N)$.

The difference between $[S(E_a,N)+S(E_b,N)]/2$ and
$S(E_a/2,N/2)+S(E_b/2,N/2)$ we call henceforth $\Delta S_{surf-corr}$. The
correct entropy balance, before and after establishing the energetic split
$e_b>e_a$ of the system, is
\begin{equation}
 S_{after}-S_{before}=\frac{S_a+S_b}{2}-\Delta S_{surf-corr.}-S_2 \le
 0 \label{balance}
\end{equation} even though the difference of the first and the last term is positive.

In the inverse direction: By {\em relaxing} the constraint and allowing, on
average, for an energy-flux ($\Delta E_{b\to a}>0$) {\em opposite to
$T_a-T_b>0$, against the temperature-gradient (slope)}, but in the
direction of the energy-slope, the entropy $S_{total}\to S_2$ increases.
This is consistent with the naive picture of an {\em energy equilibration}.
Thus {\em Clausius' "energy flows always from hot to cold", i.e. the
dominant control-role of the temperature in thermo-statistics
\cite{hertz10a} is violated}. Of course this shows again that {\em unlike
to extensive thermodynamics the temperature is not the appropriate control
parameter in non-extensive systems.}

In the thermodynamic limit $N\to\infty$ of a system with short-range
coupling $\Delta S_{surf-corr.}\sim N^{2/3}$, $\Delta
S_{surf-corr.}/N=\Delta s_{surf-corr.}\propto N^{-1/3}$ must go to $0$ due
to van Hove's theorem.

\section{The microcanonical mechanism leading to condensation, phase separation
and the origin of the convexities of $S(E)$\label{convex}} Many theoretical
applications of microcanonical thermodynamics to realistic examples of hot
nuclei, atomic clusters, and rotating astro-physical systems have been been
presented during the past 20 years and show convex intruders in the
microcanonical entropy and, consequently, negative heat capacities, c.f.
e.g. the publication list on my WEB-site http://www.hmi.de/people/gross/,
see also refs.\cite{chomaz99,chomaz00,chomaz00a,chomaz02}. Here I will
illuminate the general microscopic mechanism leading to the appearance of a
convex intruder in $S(E,N,{\cal V},\cdots)$.

I assume the system is classical and obeys the Hamiltonian:
\begin{eqnarray}H &= &\sum_i^N{\frac{p_i^2}{2m}}+\Phi^{int}[\{\nvec{r}\}]
\label{hamiltonian}\\\nonumber\\\nonumber\\
\Phi^{int}[\{\nvec{r}\}]&:=&\sum_{i<j}{\phi(\nvec{r}_i-\nvec{r}_j)}\nonumber
\end{eqnarray} In this case the system is
controlled by energy and volume.
\subsection{Liquid--gas transition\label{liquidgas}} The microcanonical sum of states or partition sum is:
\begin{widetext}
\begin{eqnarray}
W(E,N,{\cal V})&=&\frac{1}{N!(2\pi\hbar)^{3N}}\int_{{\cal
V}^N}{d^{3N}\nvec{r} \int{d^{3N}\nvec{p}_i{
\epsilon_0\;\delta(E-\sum_i^N{\frac{\nvec{p}_i^2}{2m_i}-\Phi^{int}[\{\nvec{r}\}]})}}}
\nonumber\\&=&\frac{{\cal V}^N \epsilon_0(E-E_0)^{(3N-2)/2}
\prod_1^N{m_i^{3/2}}}{N!\Gamma(3N/2) (2\pi\hbar^2)^{3N/2}} \int_{{\cal
V}^N}{\frac{d^{3N}r} {{\cal
V}^N}}\left(\frac{E-\Phi^{int}[\{\nvec{r}\}]}{E-E_0}\right)^{(3N-2)/2}
\label{split}\\\nonumber\\ &=&W_{id-gas}(E-E_0,N,{\cal V})\times
W_{int}(E-E_0,N,{\cal
V})\nonumber\\\nonumber\\&=&e^{[S_{id-gas}+S_{int}]}\label{micromeg}\\&&\nonumber\\
W_{id-gas}(E,N,{\cal V})&=&\frac{{\cal V}^N\epsilon_0
E^{(3N-2)/2}\prod_1^N{m_i^{3/2}}}{N!\Gamma(3N/2) (2\pi\hbar^2)^{3N/2}}\\
W_{int}(E-E_0,N,{\cal V})&=&\int_{{\cal V}^N}{\frac{d^{3N}r} {{\cal
V}^N}}\Theta(E-\Phi^{int}[\{\nvec{r}\}])
\left(1-\frac{\Phi^{int}[\{\nvec{r}\}]-E_0}{E-E_0}\right)^{(3N-2)/2}
\label{Win1}
 \end{eqnarray}
\end{widetext}
 ${\cal V}$ is the spatial volume. $E_0=\min
\Phi^{int}[\{\nvec{r}\}]$ is the energy of the ground-state of the
system.

The separation of $W(E,N,{\cal V})$ into $W_{id-gas}$ and $W_{int}$ is the
microcanonical analogue of the split of the canonical partition sum into a
kinetic part and a configuration part, in the micro, however, without
assuming any thermodynamic limit or homogeneity:
\begin{equation}
Z(T)=\frac{{\cal V}^N}{N!}\left(\frac{m
T}{2\pi\hbar^2}\right)^{3N/2}\int{\frac{d^{3N}r}{{\cal
V}^N}e^{-\frac{\Phi^{int}[\{\nvec{r}\}]}{T}}}\label{canonical}
\end{equation}

In the thermodynamic limit the order parameter of the
(homogeneous) liquid-gas transition is the density. The transition
is linked to a condensation of the system towards a larger density
controlled by pressure. For a finite system we expect the
analogue. However, here it is controlled by the constant available
system volume ${\cal V}$. At low energies the $N$ particles
condensate into a droplet with much smaller volume ${\cal
V}_{0N}$, the system does not fill the volume ${\cal V}$. $N-1$
internal coordinates are limited to ${\cal V}_{0N}$. Only the
center of mass of the droplet can move freely in ${\cal V}$
(remember we did not fix the C.M. in eq.(\ref{split}) ). The
system does not fill the $3N$-configuration space ${\cal V}^N$.
Only a stripe with width ${\cal V}_{0N}^{1/3}$ in $3(N-1)$
dimensions of the total $3N$-dim space and with the width ${\cal
V}^{1/3}$ in the remaining $3$ dimensions of center of mass motion
is populated. {\em The system is non-homogeneous even though it is
equilibrized and, at low energies, internally in the single liquid
phase}. It is {\em not} characterized by an intensive homogeneous
density. In fact, $W_{int}(E-E_0,N,{\cal V})$ can be written as:
\begin{eqnarray}
W_{int}(E-E_0,N,{\cal V})&=& \left[\frac{{\cal V}(E,N)}{\cal
V}\right]^N\le 1 \label{Win1b}\\ \left[{\cal
V}(E,N)\right]^N&\stdef&\nonumber\\
 \lefteqn{\hspace{-3cm}\int_{{\cal
V}^N}d^{3N}r\;\Theta(E-\Phi^{int}[\{\nvec{r}\}])}\nonumber\\
\lefteqn{\hspace{-3cm}\times\left(1-\frac{\Phi^{int}[\{\nvec{r}\}]-E_0}{E-E_0}\right)^{(3N-2)/2}}
\label{Sin2}\\\nonumber\\ S_{int}(E-E_0,N,{\cal V})&=&N\ln\left[\frac{{\cal
V}(E,N)}{\cal V}\right]\le0\label{Sin1}
\end{eqnarray}
The first factor $\Theta(E-\Phi^{int}[\{\nvec{r}\}])$ in eq(\ref{Sin2})
eliminates the energetically forbidden regions. Only the potential holes
(clusters) in the $3N$-dim potential surface $\Phi^{int}[\{r\}]$ remain.
Their volume ${\cal V}^N(E,N)\le{\cal{V}}^N$ is the accessible part of the
$3N$-dim-spatial volume where $\Phi^{int}[\{r\}]\le E$. I.e. ${\cal
V}^N(E,N)$ is the total $3N$-dim. eigen-volume of the condensate
(droplets), with $N$ particles at the given energy, summed over all
possible partitions, clusterings, in $3N$-configuration space. The relative
volume fraction of each partition compared with ${\cal V}^N(E,N)$ gives its
relative probability. ${\cal V}^N(E,N)$ has the limiting values:
\begin{equation}
{[{\cal V}(E,N)]^N}=\left\{\begin{array}{ll}{\cal V}^N&\mbox{for $E$ in the
gas phase}\nonumber\\{{\cal V}_{0N}}^{N-1}{\cal V}&\mbox{for~~}E=E_0
\end{array}\right.\label{volume}\end{equation}
$W_{int}(E-E_0,N,{\cal V})$ and $S_{int}(E-E_0,N,{\cal V})$ have
the limiting values:
\begin{eqnarray} W_{int}(E-E_0)&\le& 1,\;\Rightarrow S_{int}(E-E_0,N)
 \le0\nonumber\\
&\rightarrow& \left\{\begin{array}{ll}1 &\;\;\; \;\;\;\;\;\; \;\;\;E\gg
\Phi^{int}\nonumber\\\left[\frac{{\cal V}_{0N}}{\cal
V}\right]^{(N-1)}&\;\;\; \;\;\;\;\;\; \;\;\;E\to E_0\end{array}\right.
\\\\\nonumber\\ S_{int}(E-E_0)&\to
&\left\{\begin{array}{ll}0&E\gg \Phi^{int}\nonumber\\ln\left\{[\frac{{\cal
V}_{0N}}{\cal V}]^{N-1}\right\}< 0&E\to E_0\end{array}\right.
\label{Sin2b}\\\end{eqnarray}

All physical details are encrypted in $W_{int}(E-E_0,N,{\cal V})$ or
$S_{int}(E-E_0,N,{\cal V})$ alias $N\ln[{\cal V}(E,N)]$, c.f.
eqs.(\ref{Win1b}--\ref{Sin2b}): If the energy is high the detailed
structure of $\Phi^{int}[\{\nvec{r}\}]$ is unimportant $W_{int}\approx 1$,
$S_{int}\approx 0$. The system behaves like an ideal gas and fills the
volume ${\cal V}$. At sufficiently low energies only the minimum of
$\Phi^{int}[\{\nvec{r}\}]$ is explored by $W_{int}(E-E_0,N,{\cal V})$. The
system is in a condensed phase, a single liquid drop, which moves freely
inside the empty larger volume ${\cal V}$, the $3(N-1)$ internal dofs are
trapped inside the {\em reduced} volume ${\cal{V}}_{0N} \ll {\cal{V}}$.

One can guess the general form of $N\ln[{\cal V}(E,N)]$: Near the
groundstate $E\goo E_0$ it must be flat $\approx(N-1)\ln[{\cal
V}_{0N}]+\ln[{\cal V}-{\cal V}_{0N}]$ because the liquid drop has some
eigen-volume ${\cal V}_{0N}$  in which {\em each} particle can move
(liquid). With rising energy $\ln[{\cal V}(E,N)]$ rises up to the point
($E_{trans}$) where it is possible that the drop fissions into two. Here an
{\em additional} new configuration opens in $3N$-dim configuration space:
Either one particle evaporates from the cluster and explores the external
volume ${\cal V}$, or the droplet fissions into two droplets and the two CM
coordinates explore the larger ${\cal V}$. This gives a sudden jump in
$S_{int}(E)$ by something like $\sim \ln\{\frac{{\cal V}-{\cal
V}_{0(N-1)}}{{\cal V}_{0(N-1)}}\}$ and similar in the second case.

Of course, this sudden opening of additional parts of the
$3N$-dim. configuration space gives also a similar jump upwards in
the total entropy
\begin{eqnarray}
S(E)&=&S_{id-gas}+S_{int}\nonumber\\ &\propto& \ln{\int_{{\cal V}^N}{
d^{3N}r\left(E-\Phi^{int}[\{\nvec{r}\}]\right)^{(3N-2)/2}}}
\end{eqnarray}
by $\sim \ln\{\frac{{\cal V}-{\cal V}_{0(N-1)}}{{\cal V}_{0(N-1)}}\}$.
Later further such "jumps" may follow. Each of these "jumps" induce a
convex upwards bending of the total entropy $S(E)$ (eq.\ref{micromeg}).
Each is connected to a bifurcation and bimodality of $e^{S(E)-E/T}$ and the
phenomenon of {\em phase-separation}.

In the conventional canonical picture for a large number of
particles this is forbidden and hidden behind the familiar
Yang-Lee singularity of the liquid to gas phase transition.

In the microcanonical ensemble this is analogue to the phenomenon
of {\em multifragmentation} in nuclear systems
\cite{gross174,gross153}. This, in contrast to the mathematical
Yang-Lee theorem, {\em physical} microscopic explanation of the
liquid to gas phase transition sheds sharp light on the {\em
physical} origin of the transition, the sudden change in the {\em
inhomogeneous} population of the $3N$-dim. configuration space.

\subsection{Solid--liquid transition}
In contrast to the liquid phase, in the crystal phase a molecule
can only move locally within its lattice cage of the size $d^3$
instead of the whole volume ${\cal V}_{0N}$ of the condensate.
I.e. in eq.(\ref{Sin2b}) instead we have
$S_{int}\to\ln\{[\frac{d^3}{{\cal V}_{0N}}]^{N-1}\}$.
\subsection{Summary of section \ref{convex}}
The gas--liquid transition is linked to the transition from uniform filling
of the container volume ${\cal V}$ by the gas to the {\em smaller
eigen-volume} of the system ${\cal V}_0$ in its condensed phase where the
system is {\em non-homogeneous} (some liquid drops inside the larger {\em
empty} volume ${\cal V}$). First $3(N-1)$, later at higher energies less
and less positional degrees of freedom condensate into the drop. First
three, then more and more dofs. (C.M.-coordinates of the drops) explore the
larger container volume ${\cal V}$ leading to upwards jumps (convexities)
of $S_{int}(E)$. The volume of the container controls how close one is to
the critical end-point of the transition, where phase-separation
disappears. Towards the critical end-point, i.e. with smaller ${\cal V}$
the jumps $\ln[{\cal V}-{\cal V}_0]-\ln[{\cal V}_0]$ become smaller and
smaller. In the case of the solid--liquid transition, however, the external
volume ${\cal V}$ of the container  confines only the C.M.-position of the
crystal resp. the droplet. As long as both $d^3$ and ${\cal V}_{0N}$ remain
small compared to ${\cal V}$, the latter has no direct influence on
$S_{int}(E)$. The entropy jumps during melting are by $\Delta
S_{int}\propto\ln{{\cal V}_{0N}}-\ln{d^3}$.

At the surface of a drop $ \Phi^{int}> E_0=\min \Phi^{int}$, i.e.
the surface gives a {\em negative} contribution to $S_{int}$ in
eq.(\ref{Sin2}) and to $S$ at energies $E\goo E_0$, as was
similarly assumed in section (\ref{chsplit}) and explicitly in
eq.(\ref{balance}).
\section{Acknowledgement} I am grateful to J. M\"oller for insistent, therefore
helpful, discussions. J.F. Kenney made numerous illuminating suggestions, I
hope he finally can agree with this version.
\\
\section{Appendix, A simple model for condensation}
Let's assume the various potential-pockets are attractive square-wells with
depths $\Phi_\lambda<0$. With the (somewhat schematic) abbreviation:
\begin{eqnarray}
I(\mathbf{K})&:=&\label{massdispersion}
\\ \lefteqn{\hspace{-4.2cm}\int_{{\cal
V}^N}{d^{3N}r}\;\Theta(E-\Phi^{int}[\{\nvec{r}\}])
\left(E-\Phi^{int}[\{\nvec{r}\}]\right)^{(3N-\mathbf{K})/2}}\nonumber
\\\nonumber\\
\lefteqn{\hspace{-4.2cm}=\sum_{\lambda=1,N}{\left<(E-\Phi_\lambda)^{(3N-\mathbf{K})/2}\right>
\left({\cal V}-\sum_{k=1}^{N-\lambda}{\cal
V}_{0k}\right)^\lambda\prod_{k=1}^{N-\lambda}{\cal V}_{0k}}}\nonumber
\end{eqnarray}
Here $\Phi_\lambda\le 0$ are different topological regions of
$\Phi^{int}[\{\nvec{r}\}]$: Single cluster ($\lambda=1$), two clusters
($\lambda=2$), several clusters or, finally, if energetically possible,
($\lambda=N$) free particles, $\Phi=0$. Depending on $\lambda$ there are
$3\lambda$ C.M. coordinates which can move over the whole volume ${\cal V}$
in contrast to the rest $3(N-\lambda)$ internal coordinates which are
limited to smaller cluster volumes ${\cal V}_{0k}$.

All particles condensed into one single cluster ($\lambda=1$) will have the
lowest potential $\Phi_1$ and consequently the largest $E-\Phi$. On the
other hand the volume factor $\sim{\cal V}/{\cal V}_{0(N-1)}\gg 1$ appears
only once. Then other terms with $\lambda>1$ may dominate if $\left({\cal
V}/{\cal V}_{0,N-\lambda}\right)^\lambda \ast(E-\Phi_\lambda)^{(3N-K)/2}>
{\cal V}/{\cal V}_{0(N-1)}\ast (E-\Phi_1)^{(3N-K)/2}$.

In terms of $I(\mathbf{K})$ the partition sum $W(E)$ and its derivatives
are expressed as:
\begin{eqnarray} W(E,N,{\cal V})&\propto&I(2)\\\nonumber\\
\frac{\partial W(E,N,{\cal V})}{\partial
E}&\propto&\frac{3N-2}{2}I(4)\\\nonumber\\ \frac{\partial^2 W(E,N,{\cal
V})}{\partial E^2}&\propto&\frac{(3N-2)(3N-4)}{4}I(6)\\\nonumber\\
\beta(E)&=&\frac{3N-2}{2}\frac{I(4)} {I(2)}\label{beta}\\\nonumber\\
\beta'(E)&=&\frac{3N-2}{2}\nonumber\\
\lefteqn{\hspace{-2cm}\times\left(\frac{3N-4}{2}\frac{I(6)}{I(2)}-
\frac{3N-2}{2}\left(\frac{I(4)}{I(2)}\right)^2\right)}\label{beta'}
\end{eqnarray}To get a convexity ($\beta'\ge 0$), we must have:
\begin{equation}
\frac{I(6)I(2)}{[I(4)]^2}>1+\frac{2}{3N-4}.\label{convexCriterium}
\end{equation}
\subsection{Some examples}
\subsubsection{A single phase}
When calculating eq.(\ref{beta}--\ref{convexCriterium}) one immediately
sees that for a {\em sharply mono-dispersed} sum (\ref{massdispersion}),
$\lambda \sim \lambda_0$ (single phase)
\begin{eqnarray}I(\mathbf{K-2})&\sim&(E-\Phi_{\lambda_0})I(\mathbf{K})\label{mono}\\\nonumber\\
\lefteqn{\hspace{-3.5cm}\mbox{Then l.h.s. of criterion
(\ref{convexCriterium}) becomes 1 and there is no}}\nonumber\\
\lefteqn{\hspace{-3.5cm}\mbox{convexity. }}\nonumber\\\nonumber\\
\beta&=&\frac{3N-2}{2}\frac{1}{E-\Phi_{\lambda_0}}\\
\beta'&=&-\frac{2}{3N-2}\beta^2
\end{eqnarray}
I.e. the heat capacity is:
\begin{equation}
Nc(E)=-\frac{\beta^2}{\beta'}=\frac{3N-2}{2}
\end{equation}equal to the ideal gas value, independently of whether the
system is condensed or not. (Remember, in
eq.(\ref{massdispersion}) we assumed for simplicity that
$\Phi_\lambda$ has a flat bottom).

\subsubsection{Bimodal distribution, phase-separation}
When the massdispersion (\ref{massdispersion}) is bimodal, when e.g. two
terms in the sum dominate, i.e. when
\begin{eqnarray}
I(\mathbf{K})&\approx&\\
\lefteqn{\hspace{-4cm}\sum_{\lambda=\lambda_1,\lambda_1}
{\left<(E-\Phi_\lambda)^{(3N-\mathbf{K})/2}\right> \left({\cal
V}-\sum_{k=1}^{N-\lambda}{\cal
V}_{0k}\right)^\lambda\prod_{k=1}^{N-\lambda}{\cal V}_{0k}}}\nonumber
\end{eqnarray}
then the different $I(\mathbf{K})$ are not anymore proportional for
different $\mathbf{K}$ as in eq.(\ref{mono}), and the curvature $\beta'$,
eq.(\ref{beta'}) can very well be positive.

I.e. as anticipated in our previous discussion in section
(\ref{convex}), the volume-factor ${\cal V}/{\cal V}_0>1$ controls
whether the system can perform a phase-transition and whether the
transition has a critical end-point and  if, where it is, i.e.
where the region of convexity disappears.
\subsubsection{Hard disks, the Alder-Wainwright Transition \cite{alder62,kenney98}}
For hard disks the potential is:
\begin{eqnarray}
d&=&|\nvec{r}-\nvec{r}'|\\
\phi(d)&=&\left\{\begin{array}{rr}\infty&d\le
d_{disk}\\0&d>d_{disk}
\end{array}\right.\\
\Phi^{int}[\{\nvec{r}\}]&=&\sum_{i> k}\phi(\nvec{r}_i-\nvec{r}_k)\\
 E_0&\equiv&0\\\nonumber\\
I(\mathbf{K})&=&E^{(3N-\mathbf{K})/2}\sum_{\lambda}( {\cal V}-{\cal
V}_{0N})^\lambda{\cal V}_{0N}^{N-\lambda}.\label{IK1}\end{eqnarray} With
${\cal V}_{0N}=Nd^3/\sqrt{2}$ the closed packing volume.

Due to the criterion (\ref{convexCriterium}) there is no bimodality vs.
energy and the heat capacity is everywhere the ideal-gas value
$Nc(E)=\frac{3N-2}{2}$.

The only interesting change of $W$ is as function of the external volume
${\cal V}$. To get some feeling for this, we make the following simplified
ansatz:\\ We assume each cluster with $n$ disks has the closed-packing
volume ${\cal V}_{0n}=\frac{n}{N}V_{0N}$. Moreover, each cluster has a {\em
constant} surface layer of volume $d$ around to separate it from other
clusters. In a configuration of $\lambda$ clusters there are $3\lambda$
positional degrees of freedom that can move free in volume of ${\cal
V}-{\cal V}_0- \lambda d$ whereas $3(N-\lambda)$ internal positional
degrees of freedom are limited to the condensate volume ${\cal
V}_{0N}^{(N-\lambda)}$. For number of states we have after these
simplifying assumptions, cf. eq.(\ref{IK1}):
\begin{eqnarray}
W&\propto& \sum_\lambda W_{\lambda}\nonumber\\ W_\lambda&=&({\cal V}-{\cal
V}_{0N}-\lambda d)^\lambda\times({\cal V}_{0N})^{N-\lambda}
\end{eqnarray}

The form of $W_\lambda$ as function of the external volume is shown in
fig(\ref{fig.2}):
\begin{figure}[h]
\includegraphics*[bb =0 0 467 359, angle= 0, width=8 cm,
clip=true]{VOLUMEW.eps} \caption{Simplified Alder-Wainwright transition,
Normalized cluster-number ($\lambda$) distributions $W_\lambda({\cal V})$
as function of the external volume ${\cal V}$.\label{fig.2}}
\end{figure}

Evidently, $W_\lambda(V)$ is for every ${\cal V}$ a sharp mono-dispersed
function of number of clusters and shows {\em no} bimodality v.s.
$\lambda$. Consequently there is no phase-separation also in the $P({\cal
V})$ curve. This is {\em not} a phase-transition of first order.


\begin{thebibliography}{10}

\bibitem{gross174}
D.H.E. Gross.
\newblock {\em Microcanonical thermodynamics: Phase transitions in ``Small''
  systems}, volume~66 of {\em Lecture Notes in Physics}.
\newblock World Scientific, Singapore, 2001.

\bibitem{wehrl78}
Alfred Wehrl.
\newblock General properties of entropy.
\newblock {\em Rev.Mod.Phys.}, 50:221--260, 1978.

\bibitem{neumann27}
J.~von Neumann.
\newblock {\em G\"ott.Nachr.}, 273:20--25, 1927.

\bibitem{gross207}
D.H.E. Gross.
\newblock A new thermodynamics from nuclei to stars.
\newblock {\em Entropy}, 6:158--179, (2004).

\bibitem{kilpatrick67}
J.E. Kilpatrick.
\newblock Classical thermostatistics.
\newblock In H.~Eyring, editor, {\em Statistical Mechanics}, number~II,
  chapter~1, pages 1--52. Academic Press, New York, 1967.

\bibitem{gross183}
D.H.E. Gross.
\newblock Ensemble probabilistic equilibrium and non-equilibrium thermodynamics
  without the thermodynamic limit.
\newblock In Andrei Khrennikov, editor, {\em Foundations of Probability and
  Physics}, number XIII in PQ-QP: Quantum Probability, White Noise Analysis,
  pages 131--146, Boston, October 2001. ACM, World Scientific.

\bibitem{hertz10a}
P.~Hertz.
\newblock \"Uber die mechanische Begr\"undung der Thermodynamik II.
\newblock {\em Ann. Phys. (Leipzig)}, 33:537, 1910.

\bibitem{berdichevsky97}
Victor~L. Berdichevski.
\newblock {\em Thermodynamics of chaos and order}.
\newblock Longman, Edinburgh Gate, Harlow, England, 1997.

\bibitem{muenster69}
Arnold M\"unster.
\newblock {\em Statistical Thermodynamics}.
\newblock Springer, {1969}.

\bibitem{rugh01}
H.H. Rugh.
\newblock Microthermodynamic formalism.
\newblock {\em Phys.Rev.E}, 64:055101 1--4, 2001.

\bibitem{gross150}
D.H.E. Gross, A.~Ecker, and X.Z. Zhang.
\newblock Microcanonical thermodynamics of first order phase transitions
  studied in the potts model.
\newblock {\em Ann. Physik}, 5:446--452, 1996,and
  http://xxx.lanl.gov/abs/cond-mat/9607150.

\bibitem{chomaz99}
Ph. Chomaz and F.~Gulminelli.
\newblock Energy correlations as thermodynamical signal for phase transitions
  in finite systems.
\newblock {\em Nucl. Phys. A}, 647:153--171, 1999.

\bibitem{chomaz00}
Ph. Chomaz, V.Duflot, and F.~Gulminelli.
\newblock Caloric curves and energy fluctuations in the microcanonical
  liquid--gas phase transition.
\newblock {\em preprint}, pages 1--4, 2000.

\bibitem{chomaz00a}
Ph.Chomaz, F.Gulminelli, and V.Duflot.
\newblock Topology of event distribution as a generalized definition of phase
  transitions in finite systems.
\newblock {\em preprint}, cond-mat/0010365:1--4, 2000.

\bibitem{chomaz02}
Ph.Chomaz and F.Gulminelli.
\newblock First order phase transitions: equivalence between bimodalities and
  the Yang-Lee theorem.
\newblock {\em preprint}, cond-mat/0210456:1--10, 2002.

\bibitem{gross153}
D.H.E. Gross.
\newblock Microcanonical thermodynamics and statistical fragmentation of
  dissipative systems --- the topological structure of the n-body phase space.
\newblock {\em Physics Reports}, 279:119--202, 1997.

\bibitem{alder62}
B.~J. Alder and T.E. Wainwright.
\newblock Phase transition in elastic disks.
\newblock {\em Phys.Rev.}, 127:359--361, 1962.

\bibitem{kenney98}
J.F. Kenney.
\newblock The evolution of multicomponent systems at high pressures: I. the
  high-pressure, supercritical, gas--liquid phase transition.
\newblock {\em Fluid Phase Equilibria}, 148:21--147, 1998.

\bibitem{moretto02}
L.~G. Moretto, J.~B. Elliot, L.~Phair, , and G.~J. Wozniak.
\newblock Negative heat capacities and first order phase transitions in nuclei.
\newblock {\em Phys. Rev.C}, 66:041601(R),nucl--theor/0208024, 2002.

\bibitem{gross95}
D.H.E. Gross.
\newblock Statistical decay of very hot nuclei, the production of large
  clusters.
\newblock {\em Rep.Progr.Phys.}, 53:605--658, 1990.

\bibitem{dAgostino00}
M.~D'Agostino, F.~Gulminelli, Ph. Chomaz, M.~Bruno, F.~Cannata,
R.~Bougault,
  F.~Gramegna, I.~Iori, N.~le~Neindre, G.V. Margagliotti, A.~Moroni, and
  G.~Vannini.
\newblock Negative heat capacity in the critical region of nuclear
  fragmentation: an experimental evidence of the liquid-gas phase transition.
\newblock {\em Phys.Lett.B}, 473:219--225, 2000.

\bibitem{borderie02}
B.~Borderie.
\newblock Dynamics and thermodynamics of the liquid-gas phase transition in hot
  nuclei studied with the indra array.
\newblock {\em J.Phys.G: Nucl.Part.Phys.}, 28:R212--R227, 2002.

\bibitem{schmidt01}
M.~Schmidt, R.~Kusche, T.~Hippler, J.~Donges, W.~Kornm\"uller, B.~von
  Issendorff, and H.~Haberland.
\newblock Negative heat capacity for a cluster of 147 sodium stoms.
\newblock {\em Phys.Rev.Lett.}, 86:1191--1194, 2001.

\bibitem{schmidt02}
M.~Schmidt and H.~Haberland.
\newblock Caloric curve across the liquid-to-gas change for sodium clusters.
\newblock {\em Pys.Rev.Lett.}, 87:203402--1--4, 2001.

\bibitem{gross196}
D.H.E. Gross.
\newblock Comment on "negative heat and first order phase transitions in
  nuclei" by Moretto et al.
\newblock pages http://arXiv.org/abs/nucl--th/0208034, 2002).

\bibitem{gulminelli03}
F.~Gulminelli and Ph. Chomaz.
\newblock Comment on "negative heat capacities and first order phase
  transitions in nuclei" by l.g. Moretto et al.
\newblock {\em Caen preprint, LPCC 99-17}, pages nucl--th/0304058, 2003.

\end{thebibliography}

\end{document}